\documentclass[aps,pre,twocolumn,preprintnumbers,amsmath,amssymb]{revtex4}

\usepackage{graphics}      
\usepackage{graphicx}      
\usepackage{url}           
\usepackage{bm}            
\usepackage{amsmath}

\usepackage{fancyhdr}
\begin{document}
\title {Nonlinear normal modes in the $\beta$-Fermi-Pasta-Ulam-Tsingou chain}
\author {Nathaniel J. Fuller}
\email[]{nfuller2@buffalo.edu}
\author{Surajit Sen}
\email[]{sen@buffalo.edu}
\affiliation {Department of Physics, University at Buffalo - SUNY, Buffalo, New York 14260-1500}
\date{\today}

\begin{abstract}
Nonlinear normal mode solutions of the $\beta$-FPUT chain with fixed boundaries are presented in terms of the Jacobi sn function. Exact solutions for the two particle chain are found for arbitrary linear and nonlinear coupling strengths. Solutions for the N-body chain are found for purely nonlinear couplings. Three distinct solution types presented: a linear analogue, a chaotic amplitude mapping, and a localized nonlinear mode. The relaxation of perturbed modes are also explored using $l_{1}$-regularized least squares regression to estimate the free energy functional near the nonlinear normal mode solution. The perturbed modes are observed to decay sigmoidally towards a quasi-equilibrium state and a logarithmic relationship between the perturbation strength and mode lifetime is found.
\end{abstract}

\maketitle

\section{Introduction}
Systems of coupled nonlinear oscillators are known to be capable of collective oscillations analogous to the modes in a linear system, termed \textit{nonlinear normal modes} (NNMs) \cite{rosenberg1964}, which can display behavior that cannot be modeled by an extension of linear theory \cite{vakakis1997}. Much of the difficulty in studying the behavior of nonlinear systems arises from their non-integrability; therefore requiring numerical integration to find a solution to the equations of motion. Although a numerical solution is useful for solving a system's dynamics given a preconceived set of parameters, it is generally not obvious how the system parameters govern the form of the dynamics and whether there exists combinations of tuned parameters leading to novel solutions. This is where attaining an analytical solution for the equations of motion is most beneficial.
\par
There is already significant literature contrasting the unique behavior of NNMs with linear normal modes \cite{rosenberg1966,rand1974,vakakis1991,vakakis1992,vakakis1997,shaw1991,shaw1993} as well as their analytic descriptions in discrete systems \cite{rosenberg1960,rosenberg1962,rosenberg1964,rand1971-1,rand1971-2,sonone2014,mahan2005}. In \cite{rosenberg1962}, the form of a recursive NNM solution was discussed for a power law potential, but the exact solution was not developed. In \cite{sonone2014}, five non-trivial NNM solutions where found for the $\beta$-Fermi-Pasta-Ulam-Tsingou ($\beta$-FPUT) system in terms of the Jacobi cn function. Here, we will extend the set of Jacobi function NNM solutions for the $\beta$-FPUT chain and characterize their relaxation towards the quasi-equilibrium state \cite{sen2004,mohan2005}. These solutions showcase interesting configurations of the $\beta$-FPUT system in both their shape and infinite relaxation time when unperturbed. Furthermore, knowledge of integrable solutions such as NNMs and the relaxation of their perturbations towards equilibrium may be relevant to developing a complete understanding of the dynamics of the FPUT chain.
\par
The equation of motion for the $i$th oscillator in the $\beta$-FPUT system is
\begin{align}
\begin{split}
m_{i}\ddot{x_{i}}=\beta_{i+1}\left(x_{i+1}-x_{i}\right)^{3}+\beta_{i}\left(x_{i-1}-x_{i}\right)^{3}\\
+k_{i+1}\left(x_{i+1}-x_{i}\right)+k_{i}\left(x_{i-1}-x_{i}\right).
\label{fput}
\end{split}
\end{align}
The parameters $k$ and $\beta$ are the linear and nonlinear spring constants respectively, $m_{i}$ is the mass of the $i$th oscillator, and $x_{i}$ is the displacement from equilibrium of the $i$th oscillator. Throughout this work, we will focus on the modes of the $\beta$-FPUT system with fixed boundaries.
\par
\section{The Dual Mode $\beta$-FPUT System}
When searching for solutions of Eq. (\ref{fput}), it is natural to begin with a system having the fewest number of coupled differential equations possible. In particular, the two mass system or \textit{dual mode system} is constrained by just two simultaneous differential equations and represents a good starting point for examining Eq. (\ref{fput}). To find the NNMs supported by the dual mode system, we note that \textit{similar} NNM solutions cause the equations of motion to decouple \cite{rosenberg1964}. Since the one particle equivalent of the symmetric $\beta$-FPUT system is simply a Duffing oscillator, we assume the free response of the Duffing equation as the solution: \cite{starossek2016}
\begin{equation}
x_{i}(t)=A_{i}\;\text{sn}\left(\mu t\;|\;\kappa\right).
\label{duff}
\end{equation}
Here, $\text{sn}\left(z\;|\;\kappa\right)$ is one of the Jacobi elliptic functions with parameter $\kappa$ (not to be confused with the modulus $\sqrt{\kappa}$). $A_{i}$ is the amplitude $i$th oscillator's displacement and the frequency parameter $\mu$ is related to the oscillation period $T$ by $T=4K(\kappa)/\mu$ \cite{starossek2016} where $K(\kappa)$ is the complete elliptic integral of the first kind.
\par
Note that the second derivative of Eq. (\ref{duff}) can be put in the form $\ddot{x}_{i}=-A_{i}\mu^{2}\left(\kappa+1\right)\text{sn}\left(\mu t\;|\;\kappa\right)+2A_{i}\mu^{2}\kappa\;\text{sn}^{3}\left(\mu t\;|\;\kappa\right)$  \cite{starossek2016}. Substituting the expressions for $x_{i}$ and $\ddot{x}_{i}$ into the equations of motion leads to the following algebraic system of equations:
\begin{subequations}
\begin{align}
\label{sdm1}
2A_{1}m_{1}\mu^{2}\kappa=\beta_{2}\left(A_{2}-A_{1}\right)^{3}-\beta_{1}A_{1}^{3},\\
\label{sdm2}
-A_{1}m_{1}\mu^{2}\left(\kappa+1\right)=k_{2}\left(A_{2}-A_{1}\right)-k_{1}A_{1},\\
\label{sdm3}
2A_{2}m_{2}\mu^{2}\kappa=\beta_{2}\left(A_{1}-A_{2}\right)^{3}-\beta_{3}A_{2}^{3},\\
\label{sdm4}
-A_{2}m_{2}\mu^{2}\left(\kappa+1\right)=k_{2}\left(A_{1}-A_{2}\right)-k_{3}A_{2}.
\end{align}
\end{subequations}
By first eliminating the variables $\mu^{2}$ and $\kappa$, and with a bit of algebra, it is possible to show that the system of equations (\ref{sdm1}-\ref{sdm4}) yields a non-trivial solution only if
\begin{equation}
\label{sscon}
\beta_{2}\left(m_{1}+m_{2}\zeta\right)\left(\zeta-1\right)^{3}-\beta_{1}m_{2}\zeta+\beta_{3}m_{1}\zeta^{3}=0,
\end{equation}
is satisfied, where we have introduced the dimensionless parameter
\begin{align}
\begin{split}
\label{zdef}
\zeta&=-\frac{1}{2}\left[\frac{m_{1}}{m_{2}}\left(\frac{k_{3}}{k_{2}}+1\right)-\left(\frac{k_{1}}{k_{2}}+1\right)\right]\\
&\pm\frac{1}{2}\left\{4\frac{m_{1}}{m_{2}}+\left[\frac{m_{1}}{m_{2}}\left(\frac{k_{3}}{k_{2}}+1\right)-\left(\frac{k_{1}}{k_{2}}+1\right)\right]^{2}\right\}^{1/2}.
\end{split}
\end{align}
\par
Having a constraint on the parameters of the system is not surprising since similar normal modes are only expected for the \textit{tuned} system \cite{vakakis1991,vakakis1997}. Assuming Eq. (\ref{sscon}) is satisfied, the amplitudes of the two oscillators follow the relation
\begin{equation}
\label{sdm_a}
A_{2}=\zeta A_{1},
\end{equation}
while the frequency and parameter in terms of $A_{1}$ are
\begin{align}
\label{sdm_mu}
\mu&=\left(\frac{2\left[k_{1}-k_{2}\left(\zeta-1\right)\right]+A_{1}^{2}\left[\beta_{1}-\beta_{2}\left(\zeta-1\right)^{3}\right]}{2m_{1}}\right)^{1/2},\\
\label{sdm_k}
\kappa&=-\left(\frac{2\left[k_{1}-k_{2}\left(\zeta-1\right)\right]}{A_{1}^{2}\left[\beta_{1}-\beta_{2}\left(\zeta-1\right)^{3}\right]}+1\right)^{-1}.
\end{align}
Thus, given a particular amplitude of the first oscillator $A_{1}$, the amplitude of the second oscillator $A_{2}$, the frequency $\mu$, and the parameter $\kappa$ can be immediately computed.
\par
To illustrate the solution of a two particle NNM, we consider a system where  $m_{1}=m_{2}=m$ and $\beta_{1}=\beta_{2}=\beta_{3}=\beta$. Under these constraints, the only solutions of Eq. (\ref{sscon}) are $\zeta=\pm1$. Then, if $k_{1}=k_{2}=k$, the only positive solution of $k_{3}$ allowed by Eq. (\ref{zdef}) is $k_{3}=k$, leading to a homogeneous set of masses and spring constants. This has been predicted previously using a matching criterion for the tuned system \cite{vakakis1991,vakakis1997}. However, Eq. (\ref{sscon}) provides a more generalized tuning criterion, allowing for the calculation of NNM with non-matching masses, amplitudes, and spring constants.
\par
According to Eq. (\ref{sdm_a}), the motion of the masses is either symmetric ($\zeta>0$) or anti-symmetric ($\zeta<0$) for the homogeneous system. The expressions for the parameters $\mu$ and $\kappa$ become more intuitive when they are expressed in terms of the total energy of the system $\mathcal{U}$. These can be written as $\mu=\left(\frac{k+\frac{1}{q}\sqrt{\frac{1}{2}\beta\mathcal{U}}}{m}\right)^{1/2}$ and $\kappa=-\left(\frac{qk}{\sqrt{\frac{1}{2}\beta\mathcal{U}}}+1\right)^{-1}$ where $q=1$ ($q=3$) represents the symmetric (anti-symmetric) case. We see that as the energy is decreased towards zero, the potential is dominated by the quadratic term as expected; evidenced by the frequency parameter approaching $\sqrt{k/m}$ and the elliptic parameter $\kappa$ approaching zero.
\par
\section{The Purely Nonlinear N-Body System}
Solutions of the form given by Eq. (\ref{duff}) can be extended to a many body system if the interaction terms are either purely linear ($\beta_{i}=0$) or purely nonlinear ($k_{i}=0$) \cite{rosenberg1962}. This constraint appears since the number of algebraic equations for a $N$ particle system grows as $2N$ while the number of free parameters grows as $N+2$. This means that for systems larger than two oscillators, the system of equations is generally inconsistent, and does not have a non-trivial solution unless a specific pattern of amplitudes is found \cite{sonone2014,mahan2005} or the value of the masses and spring constants are all precisely tuned. We therefore focus on the purely nonlinear system, sometimes called an acoustic vacuum; presenting various forms of the similar NNMs, some of which lack a proper analogue in the purely linear system.
\par
Inserting the expressions for $x_{i}$ and $\ddot{x}_{i}$ into Eq. (\ref{fput}) with $k_{i}=0$, and noting that $\kappa=-1$ since the linear term is absent, leads to the following recursive formula for the amplitudes $A_{i}$,
\begin{equation}
\label{nnm_rec}
A_{i}=A_{i-1}\left\{1-\left[\frac{2\mu^{2} m_{i-1}}{\beta_{i}A_{i-1}^{2}} +\frac{\beta_{i-1}}{\beta_{i}}\left(\frac{A_{i-2}}{A_{i-1}}-1\right)^{3}\right]^{1/3}\right\}.
\end{equation}
Since we are considering a system with fixed boundaries,
\begin{equation}
\label{nnm_rec_bound}
A_{2}=A_{1}\left[1+\left(\frac{\beta_{1}}{\beta_{2}}-\frac{2\mu^{2}m_{1}}{\beta_{2}A_{1}^{2}}\right)^{1/3}\right].
\end{equation}
If the amplitude of the first oscillator $A_{1}$ and the frequency $\mu$ are known, the amplitudes can be successively computed for the rest of the chain. In order to enforce the fixed boundary condition after the $N$th oscillator, iteration over Eq. (\ref{nnm_rec}) must lead to $A_{N+1}=0$. This condition can be satisfied by using a simple root finding algorithm to solve $A_{N+1}(A_{1},\mu)=0$ for either $A_{1}$ or $\mu$: we used the \textsc{matlab} function \texttt{fzero} for this purpose.
\par
We check the validity of the analytic solution by comparing it to the numerical result from the mass-spring chain solver \textsc{pulsedyn} \cite{Kashyap2019}. There are three particular types of NNM we wish to present. One type looks quite similar to the mode of a purely linear chain. The other two NNMs however, are unique to nonlinear systems and appear quite different from the familiar linear normal modes.
\par
When the system is made to oscillate at lower frequencies and the standing mode has a wavelength of at least several particles, the NNM appears similar to the normal mode of a linear chain. To show a specific example, we choose a system consisting of 100 oscillators and set $\beta_{i}=10$, $m_{i}=1$, and $A_{1}=1$. The aforementioned root finding algorithm is used to adjust $\mu$ with an initial guess of $\mu_{\circ}=0.8$; this converges to the closest root at $\mu\approx0.774140$.
\begin{figure}[t]
\centering
\includegraphics[width=\columnwidth]{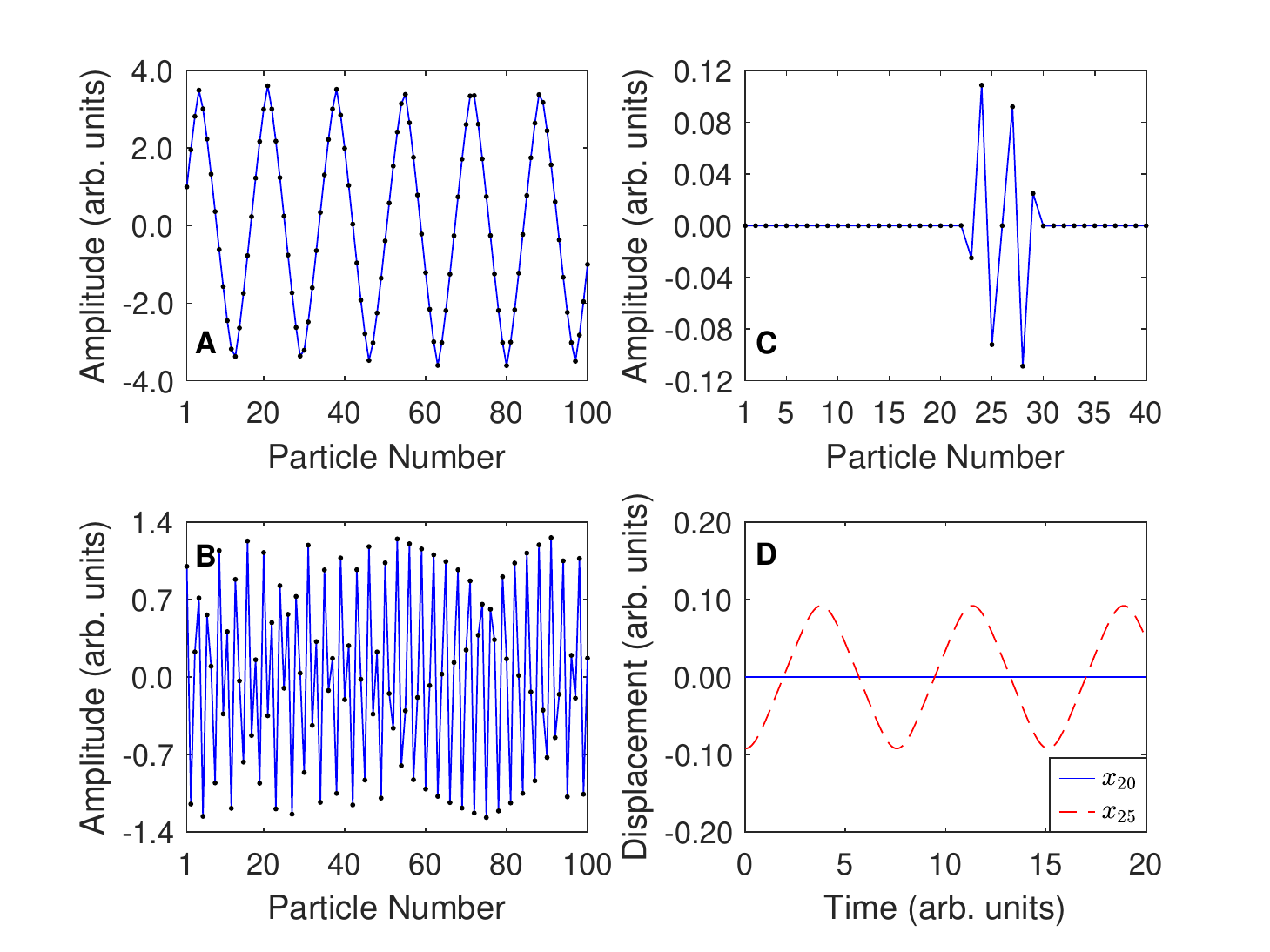}
\caption{The amplitude of each oscillator is shown for a system with \textbf{A}) $\beta_{i}=10$, $m_{i}=1$, \& $\mu\approx0.774140$, \textbf{B}) $\beta_{i}=1$, $m_{i}=1$, \& $\mu\approx2.33330$, and \textbf{C}) $\beta_{i}=10$, $m_{i}=1$, \& $\mu\approx0.694531$. Panel \textbf{D} shows the time trace of the $20$th and $25$th oscillators for the localized mode in panel \textbf{C}.}
\label{fput_am_nnm}
\end{figure}
\par
The entire set of amplitudes found from Eqs. (\ref{nnm_rec} \& \ref{nnm_rec_bound}) is plotted in panel A of Fig. \ref{fput_am_nnm}. The mode profile is reminiscent of a mode from a linear chain but is nonetheless an analytical solution of the purely nonlinear system. This is verified by comparing the analytical solution to the numerical result from \textsc{pulsedyn}. The disagreement between the two solutions is less than $1.4\times10^{-5}$ for all particle motions, indicating that the proposed solution is indeed a mode of the system. These modes are similar to those discussed in \cite{sonone2014} except that the constraints on the amplitudes are much more relaxed and the wavelength of the mode can be many particles long.
\par
When the purely nonlinear system is made to oscillate at higher frequencies, the familiar sinusoidal like modes are replaced with a set of amplitudes which chaotically oscillate. This results from the transition into chaos of the nonlinear map given by Eq. (\ref{nnm_rec}). To illustrate this, we consider a uniform chain of $100$ oscillators with $\beta_{i}=1$, $m_{i}=1$, and $A_{1}=1$. We initially guess a frequency of $\mu_{\circ}=2.4$, which converges to $\mu\approx2.33330$ using the root finding algorithm mentioned at the beginning of this section.
\par
The set of amplitudes computed using Eq. (\ref{nnm_rec}) is shown in panel B of Fig. \ref{fput_am_nnm}. To the authors' knowledge, modes of this type have not been previously shown for a mass-spring system. The set of amplitudes appear random, but have been computed almost exactly, with the only approximation entering from numerically computing $\mu$. Still, it is questionable whether such a peculiar looking mode is an admissible solution. Therefore, we again compare this to the numerical result and find the disagreement to be less than $1.5\times10^{-5}$; reassuring us of the analytical solution's validity.
\par
If the amplitude of the first oscillator in the chain $A_{1}$ is chosen to be a very small but nonzero number, the successive computations of $A_{2}$, $A_{3}$, etc. may also be comparably small. Eventually however, the nonlinear nature of Eq. (\ref{nnm_rec}) can lead to abrupt changes in the computed amplitudes by many orders of magnitude. This behavior suggest the presence of \textit{localized nonlinear modes} or \textit{breathers}.
\par
To construct a breather from Eq. (\ref{nnm_rec}), we start with a small system of six oscillators with $\beta_{i}=10$ and $m_{i}=1$. We set $A_{1}=10^{-30}$ and pick the initial guess for the frequency $\mu_{\circ}=1$, which converges to $\mu\approx0.694531$. The first two amplitudes $A_{1}=10^{-30}$ and $A_{2}\approx10^{-11}$ are exceedingly small; so the force acting on the first particle is several orders of magnitude smaller than the forces throughout the rest of the chain. Thus we suppose that energy transmission through the first particle is likewise weak, and extending the chain backwards from the first particle would not significantly change the overall solution. We also take advantage of the symmetry of Eq. (\ref{fput}) to extend the breather solution in the other direction. This gives the localized mode shown in panel C of Fig. \ref{fput_am_nnm} where we extended the system size to $40$ oscillators. The $13$ amplitudes from $A_{20}$ to $A_{32}$ are computed to construct the breather, the rest of the amplitudes in the chain are set to zero.
\par
The energy localization is illustrated in panel D of Fig. \ref{fput_am_nnm} where the motion of the $20$th oscillator, which is just outside the breather, and the $25$th oscillator, which is within the breather, is plotted from the numerical solution. The oscillatory motion of the breather remains uniform while the nearby oscillators remain at rest. We again see good agreement between the numerical and analytical results as the absolute error for any oscillator is less than $4\times10^{-7}$.
\section{Relaxation of Organized States}
The $\beta$-FPUT solutions discussed thus far are nonergodic by virtue of their integrability and hence will \textit{never} evolve to an equilibrium state. These purely nonlinear solutions are not stable however \cite{sonone2014}. Small perturbations of a NNM in a purely nonlinear system will cause the highly organized structure to eventually collapse into a collection of solitary waves, as has been similarly shown for the Korteweg-deVries equation \cite{zabusky1965}. The system will exist in a quasi-equilibrium state where the system still has not reached equipartition but has a Boltzmann like velocity distribution \cite{sen2004,mohan2005}. The system can remain in this state for possibly very long times until the secondary emissions from solitary wave collisions eventually bring the system to equipartition \cite{sen2008,sen2011}.
\par
What is less clear is how these modes relax to quasi-equilibrium. We found that perturbations of the NNMs appear to relax sigmoidally, in that the system remains organized for some time before rapidly transitioning to quasi-equilibrium. This indicates that the relaxation may be governed by relatively simple relation. In determining this relation, we note that both a particular NNM and the quasi-equilibrium state it decays into represent fixed points in the free energy landscape. The derivative of the free energy is exactly zero in a NNM configuration and nearly zero in a quasi-equilibrium state. If we can choose a proper parameter to represent the free energy functional, the relaxation dynamics of perturbed NNM can be predicted.
\par
We first define the quantity,
\begin{equation} R(t)=\left[\sum_{i=1}^{N}\left(\tilde{A_{i}^{2}}-\left<x_{i}^{2}\right>\big|_{t}\right)^{2}\right]^{1/2},
\end{equation}
representing the distance in configuration space between the mean squared configuration of a NNM and the system at time $t$. Here, $\tilde{A}_{i}^{2}$ represents the mean squared position of the $i$th oscillator of a NNM over the oscillation period $T$, and is given by $\tilde{A_{i}^{2}}=\frac{A_{i}^{2}}{T}\int_{0}^{T}dt\;\text{sn}^{2}\left(\mu t \;|-1\right)\approx 0.45695 A_{i}^{2}$. Next, we define a normalized parameter $\phi(t)=R(t)/R_{*}$ to represent the free energy functional as $F[\phi]$. $R_{*}$ is the value of $R$ in the quasi-equilibrium state. Defining $\phi(t)$ in this way gives the useful properties that it washes out oscillations on the scale of the NNMs, condenses the dynamics in N dimensions down to a single dimension with the simple interpretation of a distance in configuration space, and sets the two fixed points to $\phi(t)=0$ for the NNM and $\phi(t)=1$ for the quasi-equilibrium state.
\par
We expect the evolution of $\phi$ to be governed by the equation,
\begin{equation}
\begin{aligned}
\frac{\partial\phi(t)}{\partial t}&=-\lambda\frac{\delta F[\phi]}{\delta\phi}+\eta
\\&=-\lambda\left(a_{1}\phi+\frac{a_{2}}{2}\phi^{2}+\frac{a_{3}}{6}\phi^{3}+\cdots\right)+\eta,
\label{feeq}
\end{aligned}
\end{equation}
where $\eta$ is a noise term resulting from fluctuations not captured by $\phi$ due to the averaging in $R(t)$. Because of the simple sigmoidal relaxation character mentioned earlier, it is reasonable to assume that only a few terms dominate Eq. (\ref{feeq}) and that the noise term $\eta$ is small compared to the change in $\phi$ during the relaxation.
\par
To determine the relevant terms, we use the least absolute shrinkage and selection operator (LASSO) to perform an $l_{1}$-regularized least squares regression of data from a numerical model with the polynomial in Eq. (\ref{feeq}). This method for discovering important terms in an unknown differential equation using model data is detailed in \cite{brunton2016}. To generate the data set, we initiate a 40 particle purely nonlinear oscillator chain with $\beta=10$ and $m=1$ containing a NNM with $A_{1}=2$ and $\mu\approx0.320133$. This is a long wavelength mode with its two nodes at the boundaries of the chain. The initial positions of the particles are randomly perturbed such that the initial value of $\phi_{0}\approx7.087\times10^{-5}$.
\par
The equations of motion are solved numerically using \textsc{pulsedyn} and the value of $\phi(t)$ is calculated at each time step $t_{i}$. We use the \textsc{matlab} function \texttt{lassoglm} to perform the $l_{1}$-regularized least squares regression since it utilizes LASSO with good noise rejection. We choose the expansion in Eq. (\ref{feeq}) to contain powers up to $n=10$. Regardless of the number of terms chosen however, \texttt{lassoglm} consistently picked $a_{1}\approx-a_{2}/2$ as the only non-zero terms. The sign of $a_{1}$ must be negative since $\phi=0$ is an unstable point. Thus, the dynamics of $\phi$ can be approximately modeled by $\dot{\phi}=\lambda\left(\phi-\phi^{2}\right)$ where $a_{1}$ has been absorbed into $\lambda$. The solution of which is easily found to be
\begin{equation}
\phi(t)=\left[\left(\frac{1}{\phi_{0}}-1\right)\exp\left(-\lambda t\right)+1\right]^{-1}.
\label{prlx}
\end{equation}
\begin{figure}[t]
\centering
\includegraphics[width=\columnwidth]{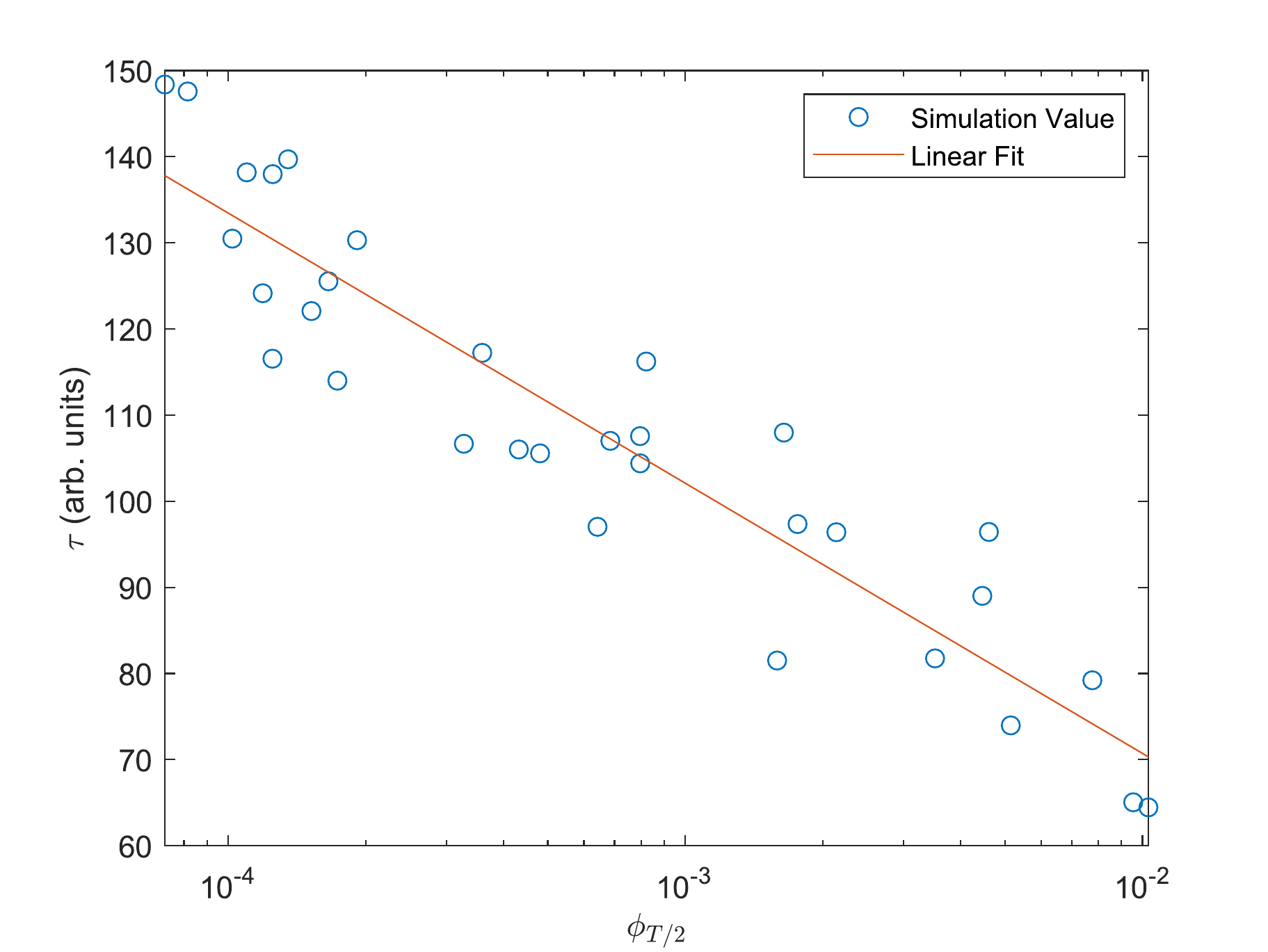}
\caption{The initial perturbation to a NNM $\phi_{T/2}$ is plotted against the mode lifetime $\tau$ on a horizontal log scale. This is done for 32 randomly generated perturbations. A linear fit is done, giving $\lambda=0.073\pm0.011$ and $b=7.0\pm15.2$.}
\label{relax_trials}
\end{figure}
\par
To determine the parameter $\lambda$ in Eq. (\ref{prlx}), we will first relate the strength of the perturbation $\phi_{0}$ to a characteristic mode lifetime $\tau$. Here, we choose the mode lifetime such that $\ddot{\phi}(\tau)$ reaches its maximum value, indicating when the NNM is most rapidly decaying. Applying this definition to Eq. (\ref{prlx}) and simplifying for $\phi_{0}\ll1$, gives the following scaling relation between $\phi_{0}$ and $\tau$:
\begin{equation}
\tau=-\frac{1}{\lambda}\ln\left(\phi_{0}\right)+b,
\label{lt}
\end{equation}
where $b$ is an undetermined constant.
\par
By running multiple simulations of the dynamics of the oscillator chain with random perturbations to the initial NNM, we can estimate the value of $\lambda$. Since $\phi_{0}$ cannot be directly calculated from the numerical simulation without negative time data, $\phi_{T/2}$ is used instead by time shifting Eq. (\ref{prlx}) by $T/2$. The value $\phi(\tau)=\frac{2-\sqrt{3}}{3-\sqrt{3}}$ is used to determine $\tau$ from the numerical integrations. The results are shown in Fig. \ref{relax_trials} where $\ln\phi_{T/2}$ is plotted against $\tau$ for 32 random perturbations. A linear fit then gives the relaxation constant $\lambda=0.073\pm0.011$.
\begin{figure}[t]
\centering
\includegraphics[width=\columnwidth]{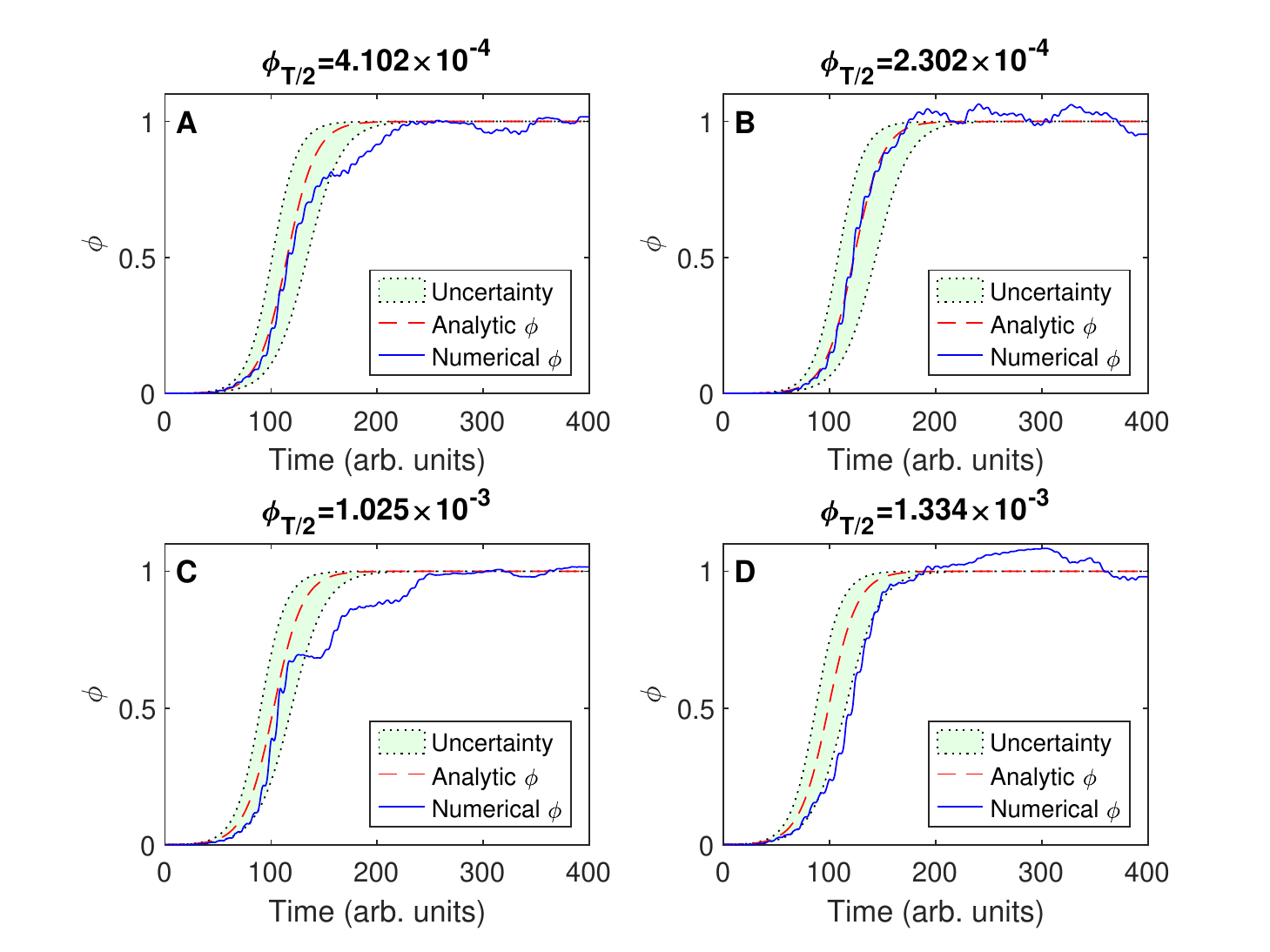}
\caption{Numerical and analytic calculations of $\phi(t)$ are shown for four random perturbations of a NNM. Panels \textbf{A}, \textbf{B}, and \textbf{C} show good agreement between Eq. (\ref{prlx}) and the numerical result, with some deviation occurring as the system approaches quasi-equilibrium. Panel \textbf{D} shows an example where the numerical solution borders the edge of the calculated uncertainty range but still follows the functional form of Eq. (\ref{prlx}).}
\label{mode_relax}
\end{figure}
\par
With the value of $\lambda$ in hand, Eq. (\ref{prlx}) can now be tested against a numerical solution. Fig. \ref{mode_relax} shows the relaxation of a NNM from four different perturbations. The predicted relaxation profile and that shown by the numerical integration agree fairly well for the four simulations shown: following the functional form of Eq. (\ref{prlx}) and only deviating as the system approaches its quasi-equilibrium state where the noise term $\eta$ becomes more important.
\par
\section{Conclusion}
Nonlinear normal mode solutions of the two particle $\beta$-FPUT system and the purely nonlinear $N$ particle system are shown in terms of the Jacobi sn function. These solutions may appear analogous to linear normal modes or display unique structures such as chaotic amplitude mappings and localized nonlinear modes. These solutions are unusual for a coupled nonlinear system, in that they are integrable and cannot achieve equipartition unless perturbed. The simple form of these solutions makes them appealing test functions for studying the relaxation dynamics of the $\beta$-FPUT system. We find that the relaxation character of the perturbed mode to a quasi-equilibrium state can be modeled with a simple sigmoid like function and that the mode lifetime is logarithmically related to the perturbation strength.

\bibliography{ref}

\end{document}